\documentclass[twocolumn]{revtex4}
\usepackage[dvips]{graphicx}
\usepackage{textcomp}
\usepackage{amsmath,amssymb}

\begin{document}

\title{Interferometric Time-Resolved Probing of Acoustic Modes in Single Gold
Nanospheres}

\author{Meindert A. van Dijk}%
\author{Markus Lippitz}
\author{Michel Orrit}
\affiliation{MoNOS, Huygens Laboratory, Universiteit Leiden, P.O.
Box 9504, 2300 RA Leiden, The Netherlands
}%

\date{18. May 2005}

\begin{abstract}

We measure the transient absorption of single gold particles with
a common-path interferometer. The prompt electronic part of the
signal provides high-contrast images for diameters as small as
10~nm. Mechanical vibrations of single particles appear on a
longer timescale (period of 16~ps for 50~nm diameter). They reveal
the full heterogeneity of the ensemble, and the intrinsic damping
of the vibration. We also observe a lower-frequency mode involving
shear. Ultra-fast pump-probe spectroscopy of individual particles
opens new insight into mechanical properties of nanometer-sized
objects.

\end{abstract}

\maketitle

The confinement of electrons and phonons causes the physical
properties of nanometer-sized objects to depart from those of bulk
solids \cite{kreibig_vollmer}. One of the ambitions of nanoscience
is to exploit these deviations, and to tailor the properties of
nanoparticles by controlling their sizes and shapes. The plasmon
resonance of noble-metal (silver and gold) particles is a
collective oscillation of the conduction electrons, which governs
their strong interaction with light. The shift and broadening of
the plasmon resonance with changes in size and shape remains an
active research area
\cite{link99oct7,hartland04physc,voisin01mar29}. A further strong
motivation for optical studies of metal nanoparticles is their
recently proposed use as labels for molecular biology
\cite{bio-gold}.

Working with ensembles of nanoparticles entails a fundamental
difficulty. The current preparation methods generate a
distribution of particles with a significant dispersion of sizes
and shapes, and with many possible configurations of defects. This
problem can be solved by the isolation of \emph{single}
nanoparticles. A number of methods have recently been put forward
to study them with far-field optical microscopy
(for a review see Ref.~\cite{vandijk05_acr}). Each single
nanoparticle being a well-defined object, it can be studied in
detail, and extended statistics can then be accumulated over many
individuals. In addition, the method gives access to environmental
influences on the particle's properties. Following one and the
same particle as a function of time unravels space- and
time-heterogeneity. In the present work, a time-resolved study of
single gold nanoparticles reveals different electronic and elastic
processes. Combining the time-resolution of short laser pulses
with the microscopy of single nanometer-sized objects (molecules
\cite{vandijk05feb25}, semiconductor structures
\cite{guenther02jul29}, or metal particles) offers new insight
into their optical and mechanical properties on their
characteristic times, picoseconds and shorter.

When an ultrashort pump pulse excites a metal particle, the
absorbed energy is first conveyed to the conduction electrons,
which collide within some tens of femtoseconds through
electron-electron interactions
\cite{link99oct7,voisin01mar29,hartland04physc}. On a 1-ps
timescale, the hot electrons thermalize with the lattice, and,
still later (typically 10~ps for a 10-nm particle), the whole
particle cools down to ambient temperature via heat diffusion. The
sudden heating of the electron gas has mechanical effects. Just as
sharp rap causes a bell to ring, an optical excitation launches
elastic oscillations, via two mechanisms: First, a short-lived
transient arises from the strong initial surge in electronic
temperature and Fermi pressure. This pressure burst is short but
strong because of the low heat capacity of the electron gas
\cite{perner00jul24}. Second, as the electronic energy is shared
with lattice modes on a picosecond timescale, anharmonicity leads
to thermal expansion. For large enough particles, both times are
short compared to the period of elastic vibrations (3.3 ps for the
breathing of a 10-nm diameter particle).

The optical properties of the hot particle are also modified. The
initial excitation spreads the electronic population around the
Fermi level, thereby opening new relaxation channels and
broadening the surface plasmon resonance \cite{perner97mar17}. The
subsequent thermal expansion of the lattice reduces the electron
density, bringing about a red shift of the plasmon resonance. Size
variations can thus be optically detected via shifts of the
plasmon resonance.

Laser-induced acoustic vibrations of nanoparticles have been
previously studied on ensembles
\cite{voisin01mar29,hartland04physc}. Such bulk observations are
only possible as far as the oscillations are synchronized. Small
differences in particle size within the ensemble lead to slight
differences in oscillation periods. This inhomogeneous broadening
often dominates the observed decay of the ensemble oscillation. Here, we
present a study of single gold nanoparticles by
interferometric pump-probe spectroscopy. Electronic and acoustic
properties are determined on a particle-by-particle basis, with
complete elimination of inhomogeneous broadening.

\begin{figure}
\includegraphics{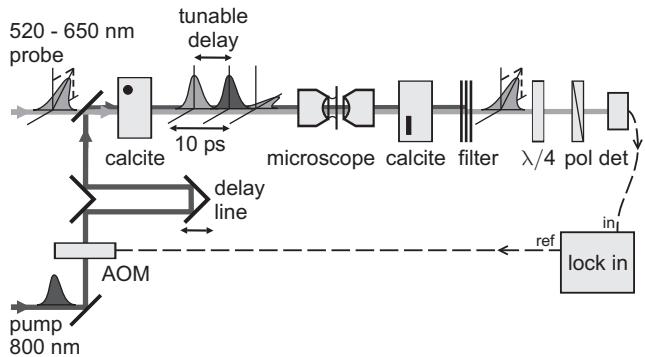}
\caption{\label{setup} Sketch of the pump-probe interferometer. A
pump pulse  and a pair of reference and probe pulses are
focused on the sample in a microscope. The reference-probe pulses
arise from a single pulse, split in time (10~ps delay) and
polarization by a properly oriented calcite crystal. The delay
between the pump pulse and the reference-probe pair can be scanned
with a delay line. After the microscope, probe and reference are
recombined by a second crystal and their interference monitors
pump-induced changes in the optical properties of the sample. A
quarter-wave plate ($\lambda/4$) and a polarizer (pol) are used to
set the working point of the interferometer. }
\end{figure}

As a single nanoparticle is much smaller than the
diffraction-limited laser spot, and as time-dependent changes in
optical properties are small, the signals are weak. Measuring
minute absorption changes requires many photons and the reduction
of all noise sources, down to photon-noise. An interferometer set
close to its dark fringe suppresses intensity fluctuations on all
timescales. We designed a common-path interferometer, in which two
interfering pulses follow the same optical path at different times
and with orthogonal polarizations (Fig.~\ref{setup}). Defects of
the optical components (particularly of the objectives), being
nearly identical for both polarizations, cancel to a large extent.

The measuring light pulse is linearly polarized at 45~degrees from
vertical and split into probe and reference pulses by a first
birefringent crystal (calcite). The reference pulse is polarized
along the crystal's horizontal fast axis, and the probe along the
vertical slow axis. The pump pulse, being polarized along one of
the crystal axes, is not split, and travels at a variable delay
from the reference-probe pulse pair.
After passage through the microscope, reference and probe pulses
are recombined in a second, identical calcite crystal, rotated so
that its fast axis is vertical. Crystals as splitting elements
have the advantage that alignment is easy, but the disadvantage
that the time delay is fixed by their thickness (here, 10~ps).

A pump-induced change in the real or imaginary part of the
particle's dielectric permittivity causes a small variation
$\Delta \zeta$ in the probe field's complex amplitude, from $E$ to
$(1+ \Delta \zeta) E$. This change is detected by the
interferometer, either as an amplitude or as a phase variation.
The working point of the interferometer is adjusted by
independently rotating a quarter-wave plate and a polarizer. The
amplitude-sensitive working point is obtained for slightly
different amplitudes but equal phases of the interfering waves.
The signal then gives the variations of the real part
$\text{Re}(\Delta \zeta)$ of the probe field. At the
phase-sensitive working point, the amplitudes are equal and the
phases are slightly different. The signal then follows
$\text{Im}(\Delta \zeta)$. In the following, unless mentioned, we
use the amplitude-sensitive working point.

The Fourier-limited 1-ps probe pulses are generated at a 76~MHz
rate by an intra-cavity, frequency-doubled, optical parametric
oscillator (OPO), tunable between 520~and 650~nm. The OPO itself
is pumped at 800~nm by the 1-ps pulses of a Ti:sapphire laser. A
small fraction of the latter beam, used as pump, is modulated at
400~kHz by an acousto-optical modulator (AOM). The probe (and
reference) power varied between 12 and 330~\textmu W, the pump
power between 0.5 and 5~mW. The home-built microscope includes an
oil-immersion objective (numerical aperture NA=1.4), and an
air-spaced objective (NA=0.95). The measurement spot has 300~nm
diameter. The sample cover slide, mounted on a piezo stage, can be
scanned with 25~nm precision. The interferometer output is fed to
an analog avalanche photodiode and demodulated in a lock-in
amplifier.

The samples are roughly spherical gold particles with diameters
ranging between 10~nm and 80~nm (British Biocell International and
Sigma-Aldrich), spin-coated with  a poly(vinyl alcohol solution
(10~g/L) on a clean glass cover slide. The polymer film was about
20~nm thick.

\begin{figure}
\includegraphics{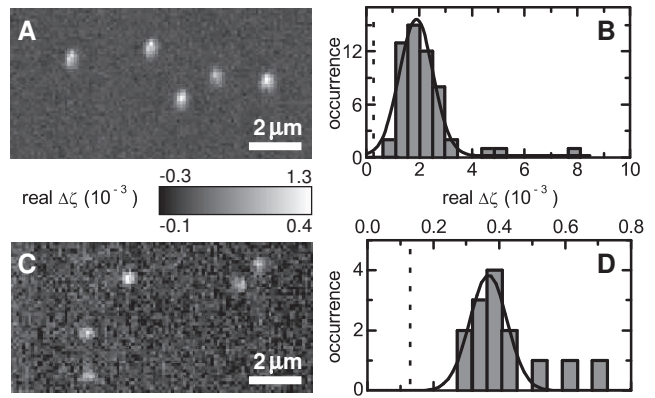}
\caption{\label{image} Raster-scanned images of single gold
nanoparticles and corresponding histograms of field changes
$\text{Re}(\Delta \zeta)$ for 20~nm (A,B) and 10~nm diameter
(C,D). The integration times were 100 and 200~ms/pixel,
respectively. A minimum signal level was required to start the fit
to a Gaussian spot, indicated by a dashed line in the histograms.
The relative width of the distributions, as deduced from rough
Gaussian fits, is 34~\% (20~nm) and 11~\% (10~nm). The noise level
in panel (C) is $7.1 \cdot 10^{-5}$, about three times the
shot-noise limit. }
\end{figure}

Figure~\ref{image} shows images of single gold nanoparticles with 10 and 20~nm
diameter. The preparation procedure was that of our earlier study
\cite{lippitz05_thg}, where third-harmonic signals proved that the particles
were isolated. Figure~\ref{image} is recorded for zero delay, i.e., when the
pump and probe pulses impinge simultaneously on the sample, providing maximum
contrast. We extracted histograms from a set of images \cite{lippitz05_thg}. The
discrimination threshold between noise and particles was $\text{Re}(\Delta
\zeta) = 2.7 \cdot 10^{-4}$~(B) and $1.3 \cdot 10^{-4}$~(D). The resulting
distributions, shown in Fig.~\ref{image}B/D, being mono-modal and well separated
from the background, confirm that each spot stems from a single particle. The
relative width of the 20-nm distribution, 34~\%, is considerably larger than
expected from the volume spread given by the manufacturer (19~\%). Additional
fluctuations in shape, orientation and surroundings of the particles can
contribute to this large dispersion via shifts and intensity changes of the
plasmon resonance, as recently observed by Berciaud et al. \cite{berciaud05mar}.
In order to minimize the spread in signal, we recorded the data of
Fig.~\ref{image} with an off-resonant probe (635~nm), which entailed a
significant signal loss. A resonant probe would directly monitor the intensity
loss of the plasmon resonance due to temperature broadening, which would provide
a stronger signal for zero delay. Shorter and resonant pulses would thus bring
the detection limit significantly below 10~nm.

\begin{figure}
\includegraphics{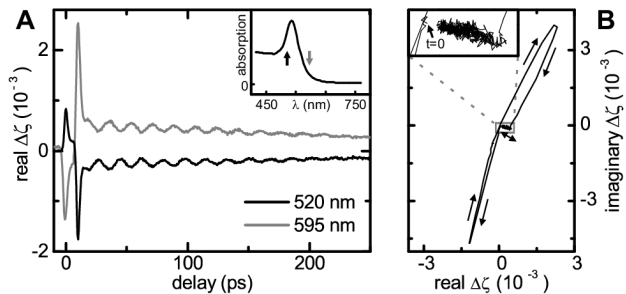}
\caption{ \label{wavelength} (A) Two delay scans of one single gold particle,
with the probe on the red side (black trace) and on the blue side (grey trace)
of the surface plasmon resonance (see inset). The period of the oscillations is
17.6~ps for both traces, corresponding to a particle size of 53~nm (assuming a
perfectly spherical particle, free boundary conditions and bulk gold properties)
\cite{hartland04physc}. (B) Full complex modification of the field by the
particle, calculated from two traces with the interferometer set to either pure
amplitude or pure phase  sensitivity ($\lambda = 520$~nm). The arrows indicate
the temporal progression.  The inset shows a 7$\times$ zoom of the center part
of the plot.}
\end{figure}

By scanning the delay between the pump and the reference-probe
pair, we monitor the time-resolved properties of the gold
particles. Figure \ref{wavelength}A shows a typical delay scan.
The prompt response ($0<t<2~\text{ps}$) stems from the broadening
of the plasmon resonance by the pump pulse. It arises when the
pump coincides with the reference and probe pulses respectively,
with opposite signs because of the interferometric difference. The
two peaks are separated by 10~ps. On a longer timescale, reference
and probe sample a damped oscillation in a differential way. From
its frequency, we assign this oscillation to the fundamental
acoustical breathing mode of a spherical particle (this purely
radial mode has no angular dependence, and no node along the
radius). Delay scans with the probe tuned to the red (595~nm) or
to the blue (520~nm) of the particle's resonance yield
out-of-phase oscillations, which clearly shows that the signal
mainly stems from a periodic shift of the plasmon resonance, as
was observed earlier in bulk experiments \cite{delfatti00may11}.

By rotating the quarter-wave plate and the polarizer, we can set
the interferometer to measure either purely absorptive or purely
dispersive effects. This attractive feature provides both the real
and imaginary parts of the a priori unknown optical response
$\Delta \zeta$, without any need for a model or for previous
knowledge. We measured delay scans for the same particle, with the
interferometer set first at the amplitude-sensitive, then at the
phase-sensitive working points. From the measured interferometer
outputs, we plotted the full $\Delta \zeta$ trace in the complex 
plane (Fig.~\ref{wavelength}B). We again
distinguish the two contributions discussed above. The fast
component due to transient electronic heating is mainly dispersive
at the chosen wavelength (520~nm). The second, slow one,
originating from acoustic vibrations and indicated with the double
arrow, is roughly perpendicular to the first part, and therefore
dominantly absorptive. The different orientations of these
components in the complex plane illustrate their different
origins, as borne out by numerical simulations.

Because of their size and shape distribution, individual particles
present slightly different oscillation periods, and they run
out-of-phase on longer timescales (see the supplementary material
for an example). This inhomogeneous broadening of the breathing
mode totally masks any intrinsic damping of the oscillation.
Selecting a single particle provides direct access to the
intrinsic (or homogeneous) damping, by removing ensemble
averaging. The damping rate is found to vary from particle to
particle, probably through fluctuations of the environment and of
the coupling to acoustic phonons in the substrate. The resonance
quality factors $Q = \nu / \Delta \nu$ ($\Delta \nu$ being the
mode's FWHM in the power spectrum), are about 4-5 for an ensemble,
but reach considerably larger values, distributed between 20 and
40, for individual particles in the thin PVA film.

\begin{figure}
\includegraphics{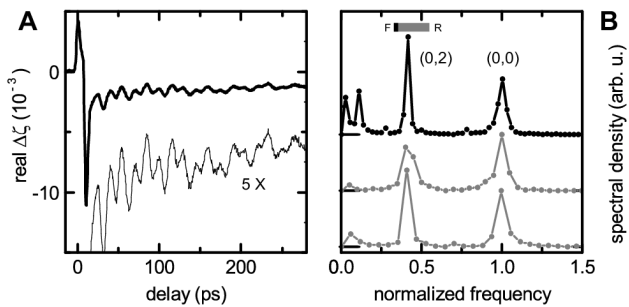}
\caption{\label{vibrmode} (A) Delay scan of a single gold nanoparticle. The
oscillation pattern shows a complex modulation.
(B) Power spectra of this particle's oscillation (top spectrum)
and of two other particles. Frequencies and amplitudes are
normalized to those of the $(0,0)$ mode (The absolute frequencies
were 67~GHz, 59~GHz, 63~GHz, from top to bottom). The low
frequency peak (top spectrum: 28~GHz) lies between calculated
frequencies of the $(0,2)$ mode for free boundary (F, thick line on
bar) and for rigid boundary (R, end of bar). The ratio of its
frequency to that of the breathing mode is constant.}
\end{figure}

The elastic vibration modes of a solid sphere are labelled in
Lamb's theory \cite{saviot96may} by two integers, $n$, the
harmonic order, i.e., the number of radial nodes, and $l$, the
angular momentum number, which represents the angular dependence
of the mode. Most of the particles only show the radial breathing
mode $(n,l)=(0,0)$ at a frequency $\Omega_{0,0}$, sometimes with a weak
trace of the higher harmonic $(n,l)=(1,0)$ at about $2.1 \cdot
\Omega_{0,0}$ \cite{saviot96may}, as  seen in ensemble measurements
\cite{nelet04mar15}. 
Some particles, however, show 
an additional mode at a lower frequency.
Fig.~\ref{vibrmode}A shows an example, where the time-response
clearly deviates from a sine-curve. The power spectrum obtained by
Fourier transformation of the oscillation (Fig.~\ref{vibrmode}B,
top trace) shows two distinct peaks. The high-frequency peak (at
67~GHz for the upper spectrum) corresponds to the spherical
breathing mode $(n,l)=(0,0)$ of a 45~nm diameter gold sphere.
Another peak appears at lower frequency (28~GHz for the upper
spectrum). This new peak cannot arise from the breathing mode of a
second, larger particle at the same spot, for the optical signal
would then be too weak (the optical response scales as the third
power of particle diameter). We assign the peak at 28~GHz to the
non-spherically symmetric $(n,l)=(0,2)$ mode, involving shear
strain (uniaxial cigar-to-pancake deformations of the sphere),
which was seen already in Raman spectra of semiconductor
nanoparticles \cite{saviot96may}. We rule out the lower-order
$(n,l)=(0,1)$ mode, corresponding to a pear-shaped deformation,
because it does not couple to the optical response, at least at
the lowest order and in a spherically symmetric environment. Both
modes can only be excited by the isotropic heat pulse if the
spherical symmetry of the particle's expansion is broken either by
the substrate, or by the particle's shape. Indeed, the
$(n,l)=(0,2)$ mode does not appear in ensemble pump-probe
experiments, where the particles' environment is isotropic. The
measured frequency of the $(n,l)=(0,2)$ mode lies between the
frequencies expected for a sphere with free and rigid boundaries
(see bar in Fig.~\ref{vibrmode}B). We note, however, that its
ratio to the breathing mode's frequency was the same for all
measured particles. The shift from the free sphere's vibration
could possibly arise from elastic perturbation by the glass
half-space.

Interferometric pump-probe measurements of single metal particles
have the double advantage of high sensitivity and of providing the
full optical response, including phase and amplitude. We have
imaged 10-nm particles with 1-ps pulses, but shorter resonant
pulses would give access to even smaller particles. Electronic
properties such as  scattering times could then be studied
on a single-particle basis. On a longer timescale, we recorded
time-traces of acoustical breathing modes, displaying
particle-to-particle fluctuations in frequency and decay rate, and
revealing the intrinsic damping of mechanical vibrations. We found
a new mode at lower frequency, so far unobserved in pump-probe
experiments, and presumably coupled by a substrate-induced
breaking of the spherical symmetry. Mechanical vibration modes are
fascinating doorways to the elastic properties of metal particles
and of their environment. More generally, probing single
nano-objects and nano-structures with short laser pulses opens a
wealth of real-time studies of nano-mechanics on picoseconds, the
characteristic vibration times at nanometer length scales.

%---------------------------------------------------------

%\onecolumn

\newpage

\begin{widetext}

% make section numbering like "S.2.4"
\renewcommand{\thesection}{\Alph{section}}
\renewcommand{\thefigure}{S.\arabic{figure}}
\setcounter{section}{18}      %18 = S
\renewcommand{\arraystretch}{1.1}

\section{Supplementary material}

% ***************************************************************************

\subsection{Experimental details of the figures}

We give here more details of the experimental conditions under which the data
presented in the figures was acquired. The particle diameter
$d_{\text{nominal}}$ is the nominal diameter of the particles, and
$d_{\text{sample}}$ the batch average diameter as given by the manufacturer. The
size dispersion, as given by the manufacturer, is in all cases about 6~\%.
$d_{\text{measured}}$ indicates the particle size as deduced from the
oscillation frequency. The  power in the visible $P_{\text{VIS}}$ is the average
power of the reference and the probe pulses together at the sample position. The
 power in the near infrared $P_{\text{NIR}}$ ($\lambda = 800$~nm) is the 
average power of the pump at the sample position, while the AOM is transmitting
(otherwise it is zero). $t_{\text{bin}}$ is the pixel dwell time.
$t_{\text{lock-in}}$ gives the time-constant of the lock-in amplifier. The
filter roll-off was always 24 dB/oct. $I_{\text{wp}} / I_0$  gives the
transmission of the interferometer at the selected working point. It is 0 at the
dark fringe and 1 at the bright fringe.

\begin{table}[h!]
\center
\begin{tabular}{l|rrrrr}
          Figure                 & 2A      &  2C     &  3      &  4  & S.1/2 \\
	  \hline
$d_{\text{nominal}}$ (nm)       & 20      &    10   & 50     &  50    & 50 \\
$d_{\text{sample}}$ (nm)       & 22      &    11   & 52     &  52    & 52 \\
$d_{\text{measured}}$ (nm)       & ---      &    ---   & 53     &  45, 48, 51    & --- \\
$\lambda_{\text{probe}}$ (nm)    & 635     &    635  & (*) &  595  & 595 \\
$P_{\text{VIS}}$  (\textmu W)  & 160     &    330  & 40      &  12   & 16  \\
$P_{\text{NIR}}$  (mW)          & 5.2     &    5.2  & 0.5    &  2.0    & 0.5 \\
$t_{\text{bin}}$  (ms)           & 100     &    200  & 470     &  100  & 160 \\
$t_{\text{lock-in}}$  (ms)       & 10      &    30   & 30      &  10   & 10 \\
$I_{\text{wp}} / I_0$          & 0.015     &    0.010 & 0.100     &  0.100  &
0.100 \\
\end{tabular}
\caption{Experimental details of the figures in the article and the
supplementary material. (*) see caption of Fig.~3.}
\end{table}

%\newpage

\subsection{Inhomogeneous broadening in an ensemble}

A striking illustration of the strength of single-particle observations is
displayed in Fig.~\ref{broadening}. The left panel shows delay scans and the
right panel power spectra of the oscillatory behavior of nanoparticles. We
measured 29 single particles under the same experimental conditions and then
averaged them to yield an "ensemble" signal (thick grey line). Examples of the
single particle data are shown as thin black lines. Comparing the "ensemble"
signal to the single particle signal, we clearly see that the ensemble
oscillation damps much faster than those of individual nanoparticles, which
broadens the peak in the ensemble power spectrum. Because of the size
distribution, the nanoparticles present slightly different oscillation periods,
and they run out-of-phase on longer timescales. This inhomogeneous broadening
easily masks the intrinsic decay of the vibrations of individual particles.

\begin{figure}[h!]
\center
\includegraphics*{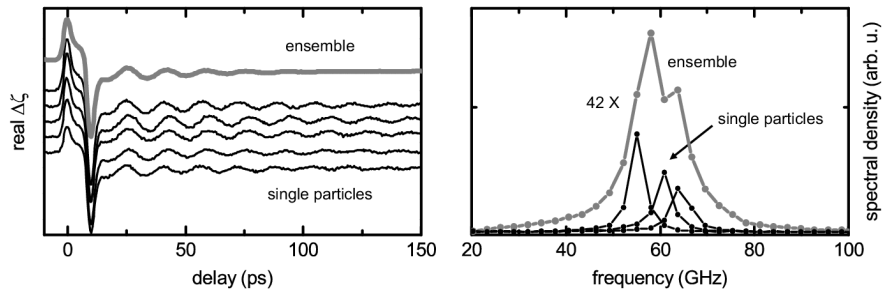}
\caption{\label{broadening} Comparison of delay scans (left panel) and power
spectra of the oscillatory part (right panel) of a set of 29 single  gold
nanoparticles (examples shown as thin black lines) to a reconstructed ensemble
measurement (thick grey line). The particles have a nominal size of 50 nm with a
spread of 6~\%. The sample was prepared as described in the article and
additionally coated with an index-matching fluid. The average spectrum in part
(B) is scaled by 42  for clarity. The quality factor $Q = \nu / \Delta \nu$ is 4
for the ensemble and about 13 for the individual traces.  The index-matching
fluid reduces the quality factor of the individual traces compared to particles 
embedded in a thin polymer film only (see main text).
}
\end{figure}

\ \\

\begin{figure}
\center
\includegraphics*{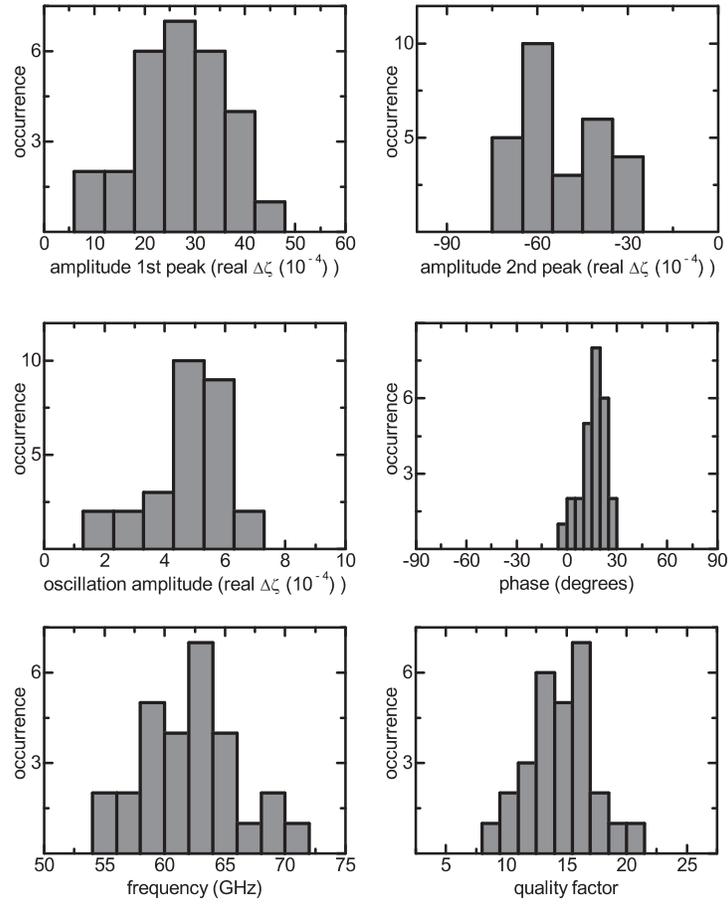}
\caption{\label{stats} Detailed statistics of the delay traces of
Fig.~\ref{broadening}. The maximum amplitude of the first ($t=0$~ps) and the
second peak ($t=10$~ps) are given. A damped cosine function was fitted to the
oscillations.  The temporal zero of this function was chosen to lie at the
second peak. The amplitude and phase  of the  damped cosine function is given in
the histograms. The oscillation frequency $\Omega_{0,0} / 2 \pi$ and quality
factor $Q$ are extracted from the Fourier-transform. The spread of the first
peak's amplitude (about 34~\%) is  larger than the expected volume spread
(19~\%). The oscillation frequency distribution on the other hand reproduces the
 size spread of about 6~\% expected from Lamb's relation $ \Omega_{0,0} \propto
v_L / d $ ($v_L$ being the longitudinal speed of sound in the sphere). 
}
\end{figure}

%--------------------------------------------------------

\newpage

\subsection{Second harmonic of the breathing mode}

\ \\

\begin{figure}[h!]
\center
\includegraphics*{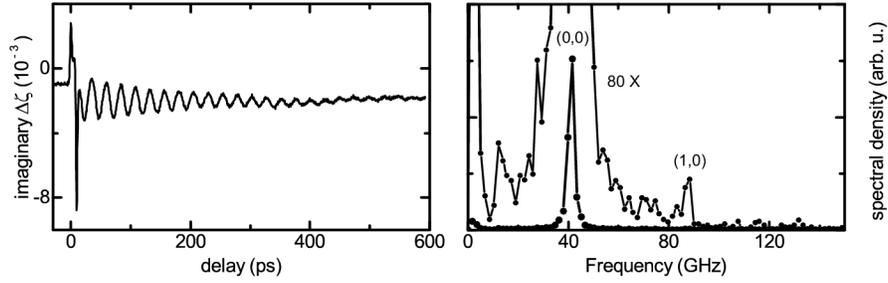}
\caption{Some particles show the second harmonic $(n,l) = (1,0)$ of the
fundamental breathing mode at $\Omega_{1,0} \approx 2.17 \cdot \Omega_{0,0}$,
close to the expected value of Lamb's theory at $2.103 \cdot \Omega_{0,0}$. The
delay trace (left panel) has no  signs of higher harmonics, but the power
spectrum (right panel) shows a clear peak at $\Omega_{1,0}$.
}
\end{figure}

%--------------------------------------------------------

\subsection{Cross-section of Figure 2C}

\ \\

\begin{figure}[h!]
\center
\includegraphics*{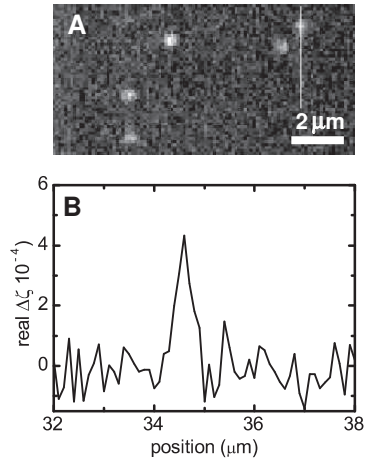}
\caption{ Cross-section through the raster-scanned image of 10-nm gold particles
shown in Fig.~2C. The noise level is $7.1 \cdot 10^{-5}$ (standard deviation),
about three times the shot-noise limit. The signal-to-noise ratio is about 6.
}
\end{figure}

\end{widetext}

\end{document}